\documentclass[final,3p,times,twocolumn]{elsarticle}
\usepackage{amssymb,amsmath,lineno}
\usepackage{pst-node,pstricks,graphicx}
\usepackage{booktabs}
\usepackage{xspace}

\usepackage{booktabs}
\usepackage{nicefrac}
\usepackage[color]{showkeys}
\usepackage{fancyhdr}

\usepackage{units}
\begin{document}                % INITIALIZE AAAzzzDD--- 
%\flushright{21.12.2005}
\begin{frontmatter}
  \title{Searching for tau neutrinos with Cherenkov telescopes}

\author[a,c]{D.~G\'ora}

\author[b]{E.~Bernardini}

\author[a]{A. Kappes}

\cortext[cor1]{E-mail address of corresponding author: Dariusz.Gora@desy.de}

\address[a]{Erlangen Centre for Astroparticle Physics,
  Friedrich-Alexander-Universit\"at Erlangen-N\"urnberg, Erwin-Rommel-Str. 1, D-91058 Erlangen,
  Germany}

\address[b]{DESY, Platanenallee 6, D-15738 Zeuthen, Germany}

\address[c]{Institute of Nuclear Physics PAN, Radzikowskiego 152, Krak\'ow, Poland}

\begin{abstract}
Cherenkov telescopes have the capability of detecting high energy tau neutrinos in the energy range of 1--1000\,PeV by searching for very inclined showers. If a tau lepton, produced by a tau neutrino, escapes from the Earth or a mountain, it will decay and initiate a shower in the air which can be detected by an air shower fluorescence or Cherenkov telescope. In this paper, we present detailed Monte Carlo simulations of corresponding event rates for the VERITAS and two proposed Cherenkov Telescope Array sites: Meteor Crater and Yavapai Ranch, which use representative AGN neutrino flux models and take into account topographic conditions of the detector sites. The calculated neutrino sensitivities depend on the observation time and the shape of the energy spectrum, but in some cases are comparable or even better than corresponding neutrino sensitivities of the IceCube detector. For VERITAS and the considered Cherenkov Telescope Array sites the expected neutrino sensitivities are up to factor 3 higher than for the MAGIC site because of the presence of surrounding mountains.
\end{abstract}
\end{frontmatter}

\section{Introduction}
Neutrinos have long been anticipated to help answering some fundamental questions in astrophysics like the mystery of the source of the cosmic rays (for a general discussion see \cite{uhecr}). For neutrinos in the TeV range, prime source candidates are Galactic supernova remnants \cite{remnats}. Neutrinos in the PeV range and above are suspected to be produced by Active Galactic Nuclei (AGN) and Gamma Ray Bursts (GRB) with many AGN models predicting a significant neutrino flux \cite{Atoyan:2004pb,PhysRevD.80.083008,Mucke2003593}. Recently, the IceCube Collaboration has reported the very first observation of a cosmic diffuse neutrino flux which lies in the 100 TeV to PeV range~\cite{science_icecube}. Individual sources, however, could not be identified up to now. While many astrophysical sources of origin have been suggested \cite{icecube_inter}, there is yet not enough information to narrow down the possibilities to any particular source.

Due to the low interaction probability of neutrinos, a large amount of matter is needed in order to detect them. One of the detection techniques is based on the observation of inclined extensive air showers (EAS) induced by taus from tau neutrino interactions deep in the atmosphere. As these showers are initiated close to the surface of the Earth, they are still very young when reaching the detector and hence have a significant electromagnetic component leading to a broad time structure of the detected signal. In contrast, showers from cosmic-ray nuclei are induced in the upper atmosphere and therefore have a strongly reduced electromagnetic component when reaching the detector. However, because of its low density, neutrino interactions are not very likely to happen inside the atmosphere. A solution to this problem is to look for so-called Earth skimming (up-going) tau neutrinos  \cite{fargion0,fargion1,zas,Letessier:2001,feng,tseng,aramo} which interact within the Earth or a mountain and produce a tau. For neutrino energies of about a EeV, the charged leptons have a range of a few kilometers and hence may emerge from the Earth or mountain, decay shortly above the ground and produce EAS detectable by a surface detector.
In some cases, two consecutive EASs might be observable, one coming from a tau neutrino interaction close to the surface and one from the decay of the resulting tau lepton. These two showers, coming from the same direction in a time interval corresponding to the tau decay time, can generate a unique signature in the detector called a Double-Bang event ~\cite{learne}. The detection of such a Double-Bang event would be very important both from the astrophysical and the particle physics point of view, as it would be an unambiguous sign for an ultra-high energy (UHE) tau-neutrino.
Up to now, there has been no clear identification of tau neutrinos at high energies.

The detection of PeV tau neutrinos through optical signals also seems possible. A combination of fluorescence and Cherenkov light detectors in the shadow of steep cliffs could achieve this goal  {\bf ~\cite{fargion0,fargion1,doi:10.1142/S0217732304014458}}. Recently, it has also been shown by the All-sky Survey High Resolution Air-shower detector (Ashra) experiment, that such kind of experiments could be sensitive to tau neutrinos from fast transient objects such as nearby GRBs~\cite{Asaoka:2012em}.We note, that the recent IceCube results do not show any neutrino events at or above the Glashow resonance at 6.3 PeV~\cite{science_icecube}. This likely means that there is either a cutoff in the astrophysical neutrino flux below $\sim 6$ PeV or the neutrino spectrum is steeper than the usually assumed $E^{-2}$ spectrum.

In principle also existing Imaging Air Cherenkov Telescopes (IACTs) such as MAGIC~\cite{magic}, VERITAS~\cite{veritas} and H.E.S.S.~\cite{hess} have the capability to detect PeV tau neutrinos by searching for very inclined showers~\cite{fargion}. In order to do that, the Cherenkov telescopes need to be pointed in the direction of the taus escaping from the Earth crust, i.e.\ at or a few degree below the horizon. This is because the trajectory of the tau lepton has to be parallel to the pointing direction of the telescope within a few degrees as the Cherenkov light is very much beamed in the forward direction. For example, the MAGIC telescope is placed on top of a mountain on La Palma at an altitude of about 2200\,m a.s.l. Thus, it can look down to the Sea and monitor a large volume within its field of view (FOV). In \cite{upgoing_magic}, the effective area for up-going tau neutrino observations with the MAGIC telescope was calculated analytically with the maximum sensitivity in the range from 100\,TeV to $\sim$~1\,EeV. However, the calculated sensitivity for diffuse neutrinos was very low because of the limited FOV, the short observation time and the low expected neutrino flux.

On the other hand, if flaring or disrupting point sources such as GRBs are observed, one can  expect an observable number of events even from a single GRB if close by. In the case of MAGIC, however, the topographic conditions allow only for a small window of about 1 degree width in zenith and azimuth to point the telescope downhill. In case of other IACT sites with different topographic conditions, the acceptance for up-going tau neutrinos will be increased by the presence of mountains. Mountains can work as an additional target and will lead to an enhancement in the flux of emerging tau leptons. A target mountain can also shield against cosmic rays and star light. 

For Cherenkov telescope sites, very often nights with high clouds prevent the observation of gamma-ray sources. In such conditions, pointing the telescopes to the horizon could significantly increase the observation time and the acceptance for up-going tau neutrinos. Next-generation Cherenkov telescopes, i.e.\ the Cherenkov Telescope Array (CTA)~\cite{cta}, can in addition exploit their much larger FOV (in extended observation mode), a higher effective area and a lower energy threshold.

In this work, we present an update of the work in~\cite{ouricrc}, where a detailed Monte Carlo simulation of event rates induced by Earth skimming tau neutrinos was performed for an ideal Cherenkov detector. Neutrino and lepton propagation was simulated taking into account the local topographic conditions at the MAGIC site and at four  possible locations of Cherenkov instruments: two in Argentina (San Antonio, El Leoncito), one in Namibia (Kuibis) and one on the Canary Islands (Tenerife). In this work, similar simulations have been performed for the location of the VERITAS telescopes and for two sites located close to VERITAS: Meteor Crater and Yavapai Ranch. These two sites were recently also considered as possible locations for CTA. Results are shown for a few representative neutrino fluxes expected from giant AGN flares. We would like to stress that in this work we are exploring the effect of different topographic conditions rather than providing a comprehensive survey of potential sites.

\section{Method}

The propagation of a given neutrino flux through the Earth and the atmosphere is simulated using  an extended version of the ANIS code~\cite{gora:2007}. For fixed neutrino energies, $10^{6}$ events are generated on top of the atmosphere with zenith angles ($\theta$) in the range $90^{\circ}$--$105^{\circ}$ (up-going showers) and with azimuth angles in the range $0^{\circ}$--$360^{\circ}$. Neutrinos are propagated along their trajectories of length $ \Delta L$ from the generation point on top of the atmosphere to the interaction volume,  defined as the volume which can contribute to the expected event rate, in steps of $\Delta L$/1000 ($\Delta L/1000 \geq 6$ km). At each step of propagation, the $\nu$--nucleon interaction probability is calculated according to  parametrization of its cross section based on the chosen parton distribution function (PDF). In particular, the propagation of tau leptons through the Earth is simulated. All computations are done using digital elevation maps (DEM)~\cite{dem} to model the surrounding mass distribution of each site under consideration. The flux of the leptons emerging from the ground as well as their energy and the decay vertex positions are calculated inside an interaction volume, modeled by a cylinder with radius of 35\,km and height 10\,km.
The detector acceptance for an initial neutrino energy $E_{\nu_\tau}$ is given by:
%\begin{equation}
\begin{eqnarray}
  A(E_{\nu_\tau})  =N_{\mathrm{gen}}^{-1} \times \sum_{i=1}^{N_{k}}
   P_{i}(E_{\nu_\tau},E_{\tau},\theta) \nonumber \\ 
   \times T_{\mathrm{eff},i}(E_{\tau},x,y,h,\theta) \times
   A_i(\theta)\times \Delta \Omega,
\label{aperture}
\end{eqnarray}
%\end{equation}
where $N_{\mathrm{gen}}$ is the number of generated neutrino events. $N_k$ is the number of $\tau$ leptons with energies $E_{\tau}$ larger than the threshold energy $E_{\mathrm{th}}=1$\,PeV and a decay vertex position inside the interaction volume. $P(E_{\nu_\tau},E_{\tau},\theta)$ is the probability that a neutrino with energy $E_{\nu_\tau}$ and zenith angle $\theta$  produces a lepton with energy $E_{\tau}$ (this probability was used as "weight" of the event). $A_i(\theta)$ is the physical cross-section of the interaction volume seen by the neutrino and $\Delta\Omega$ is the solid angle. $T_{\mathrm{eff}}(E_{\tau},x,y,h,\theta)$ is the trigger efficiency for tau-lepton induced showers with the decay vertex position at ($x$, $y$) and height $h$ above the ground. The trigger efficiency depends on the response of a given detector and is usually estimated based on Monte-Carlo simulations. In this work, we used an average trigger efficiency extracted from~\cite{Asaoka:2012em}, namely $\langle T_{\mathrm{eff}} \rangle =10$\%, which is comparable to what was calculated for up-going tau neutrino showers studied in~\cite{doi:10.1142/S0217732304014458}. This is a qualitative estimation and as such it is the major source of uncertainty on the results presented hereafter.
\begin{figure}                                                                                                             
  \begin{center}                                                                                                              
    \includegraphics[width=\columnwidth]{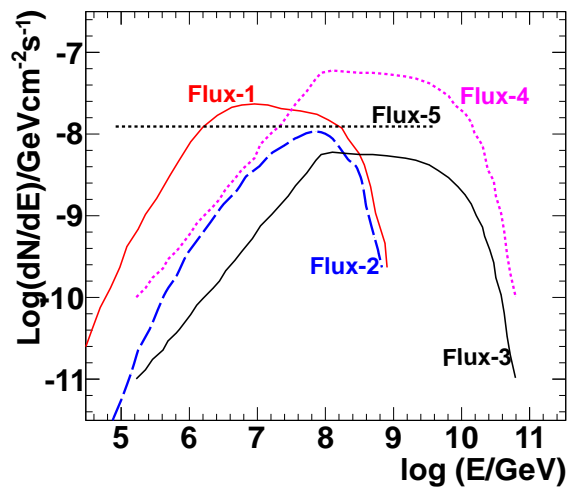}
  \end{center}                                                                                                               
  \caption{\label{fig::spectrum2} {A sample of representative neutrino fluxes from photo-hadronic interactions in AGNs. See text for more details. Flux-1 and Flux-2 are calculations for  $\gamma$-ray flare of 3C 279~\cite{2009IJMPD}. Flux-3 and Flux-4 represent predictions for PKS~2155-304~\cite{Becker2011269}. Flux-5 corresponds to a prediction for 3C~279 calculated in~\cite{PhysRevLett.87.221102}. }} 
\end{figure}   
Equation~(\ref{aperture}) gives the acceptance for diffuse neutrinos. The acceptance for a point source can be estimated as the ratio between the diffuse acceptance, defined in Eq.~(\ref{aperture}), and the solid angle covered by the diffuse analysis, multiplied by the fraction of time the source is visible $f_{\mathrm{vis}}(\delta_{s},\phi_{\mathrm{site}})$. This fraction depends on the source declination ($\delta_{s}$) and the latitude of the observation site ($\phi$). In this work, the point source acceptance is calculated as
 $ A^{\mathrm{PS}}(E_{\nu_\tau})\simeq A(E_{\nu_\tau}) / \Delta \Omega \times f_{\mathrm{vis}}(\delta_{s},\phi_{\mathrm{site}})$.
For energies above 1\,PeV, neutrinos no longer efficiently penetrate the Earth and are
preferentially observed near the horizon where they traverse a reduced chord of the Earth. Thus the main contribution to the acceptance comes from a few degrees below horizon. In this work, 
we simulated propagation of tau neutrinos with zenith angles in the range $90^{\circ} - 105^{\circ}$ corresponding to $\Delta \Omega=1.62$\,sr. 
 
\begin{figure*}[!ht]
  \begin{center}                                                                                                   
\includegraphics[width=\columnwidth,height=6.1cm]{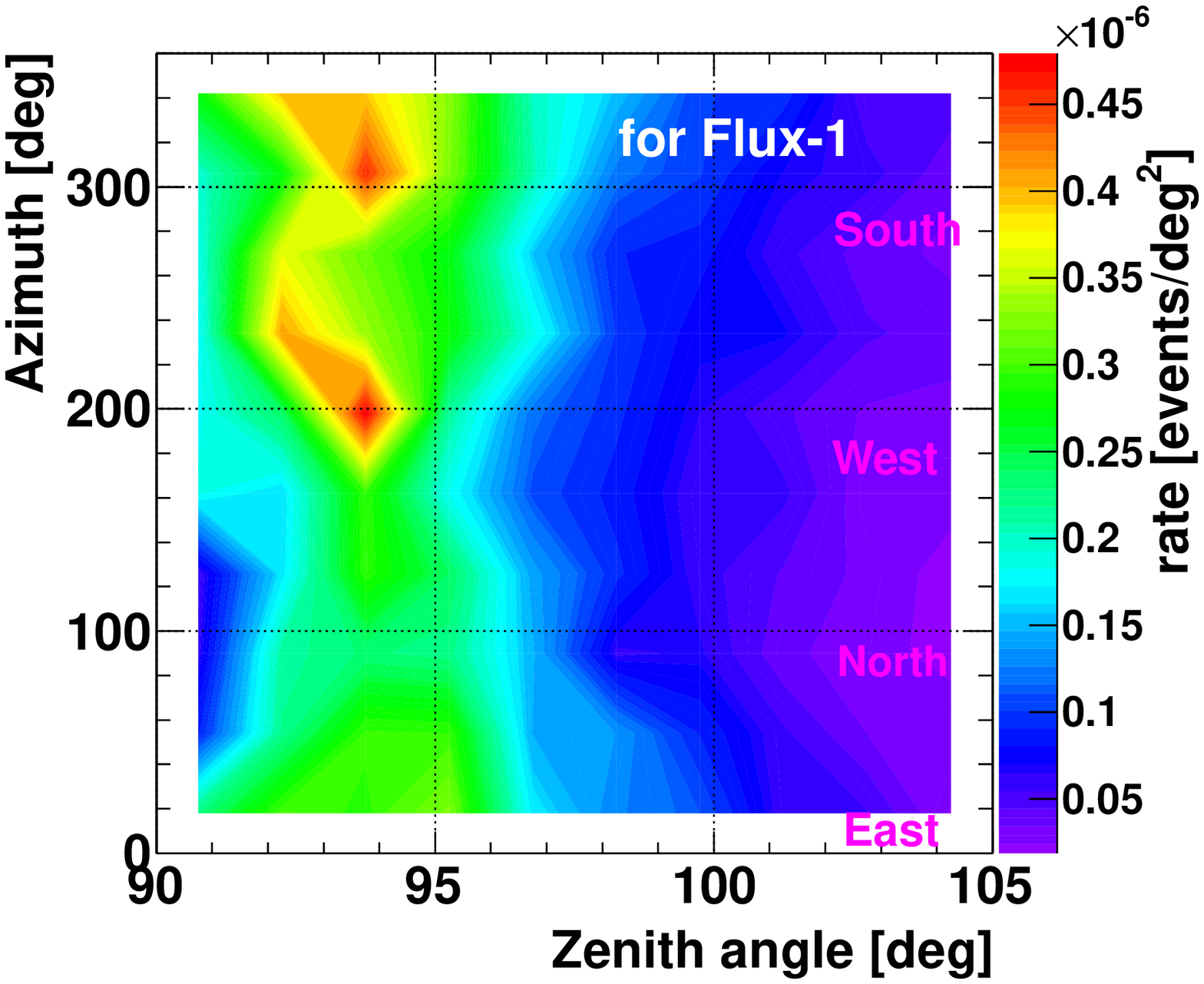} 
 \includegraphics[width=\columnwidth,height=6.1cm]{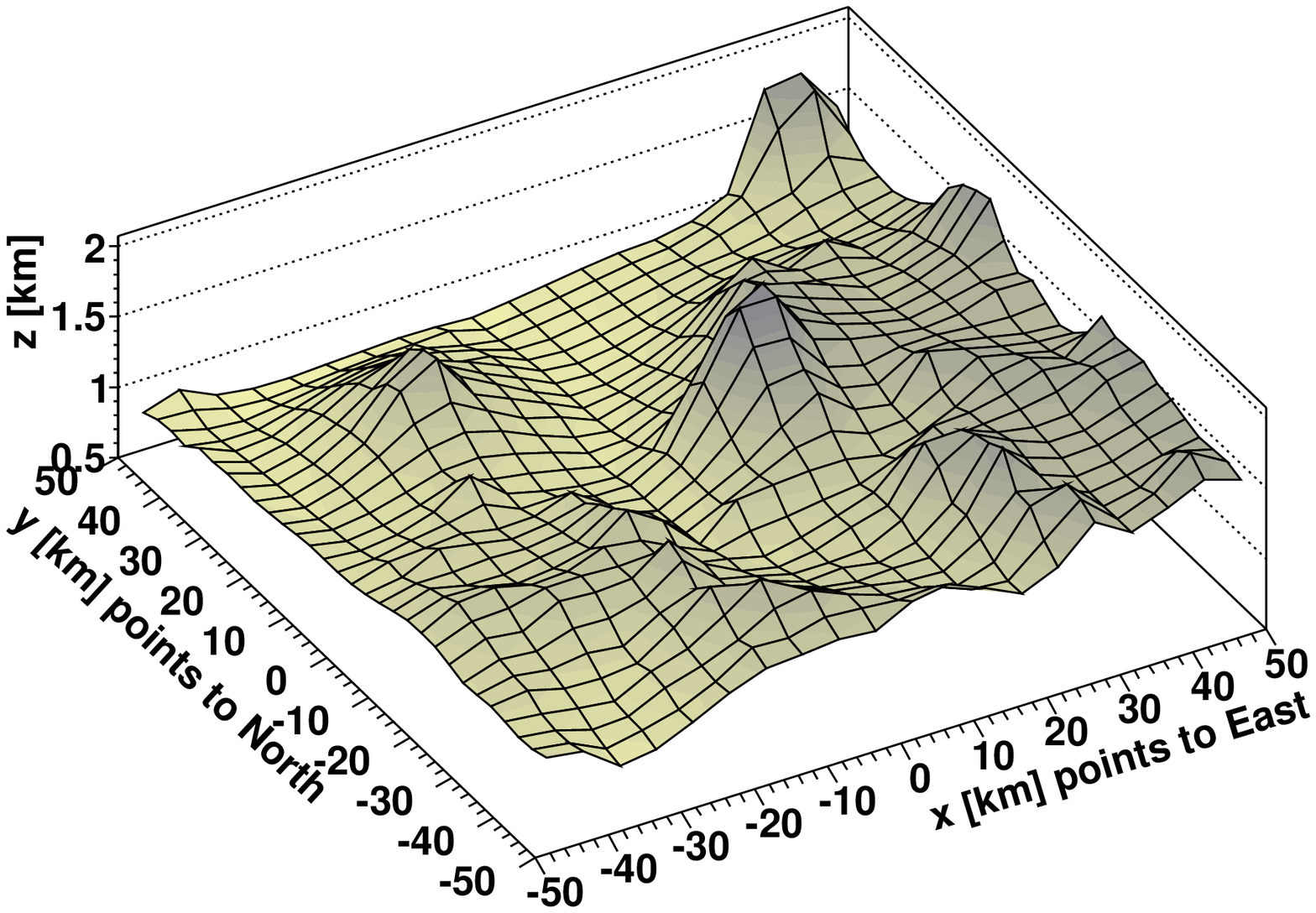}   

\includegraphics[width=\columnwidth,height=6.1cm]{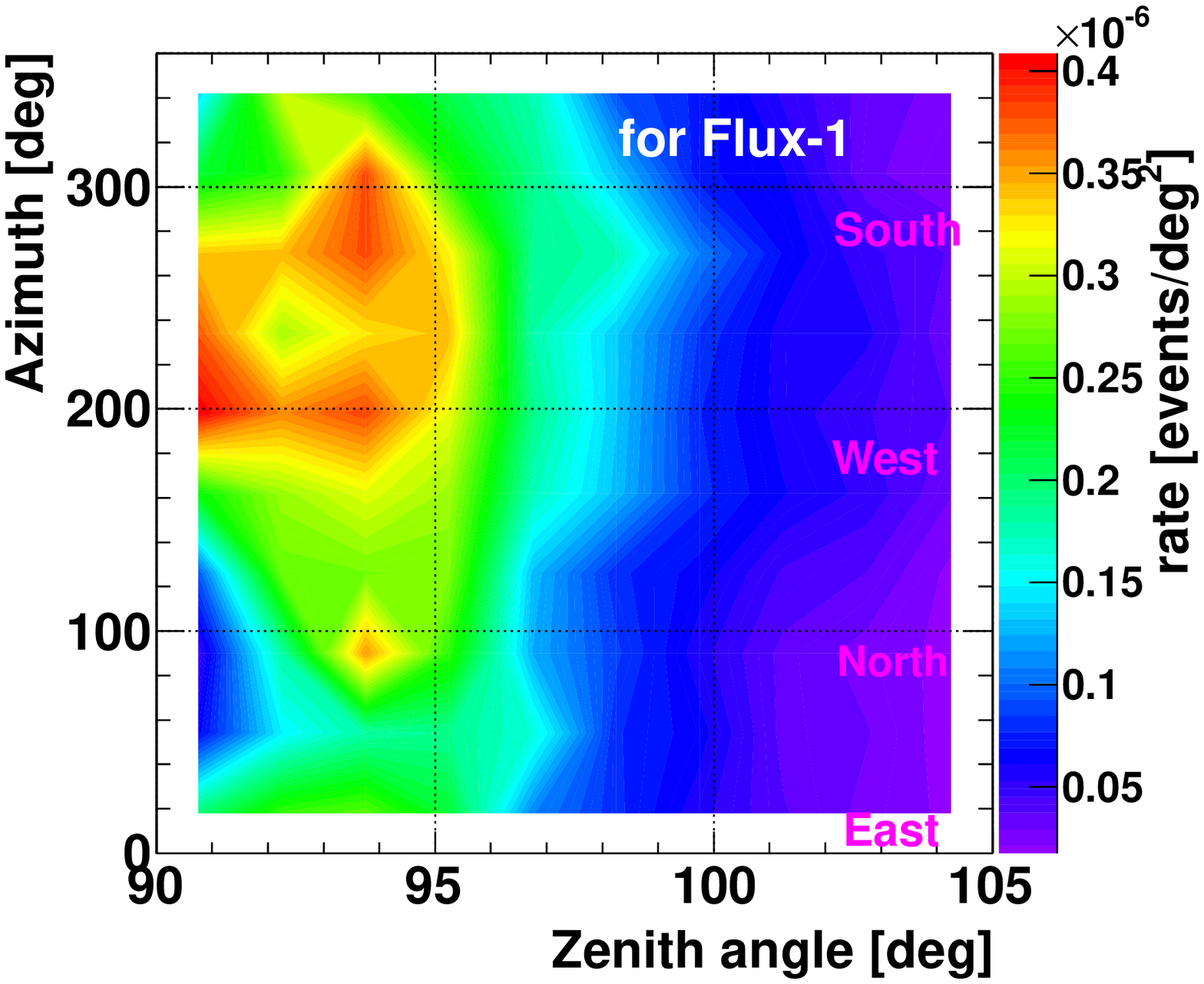} 
 \includegraphics[width=\columnwidth,height=6.1cm]{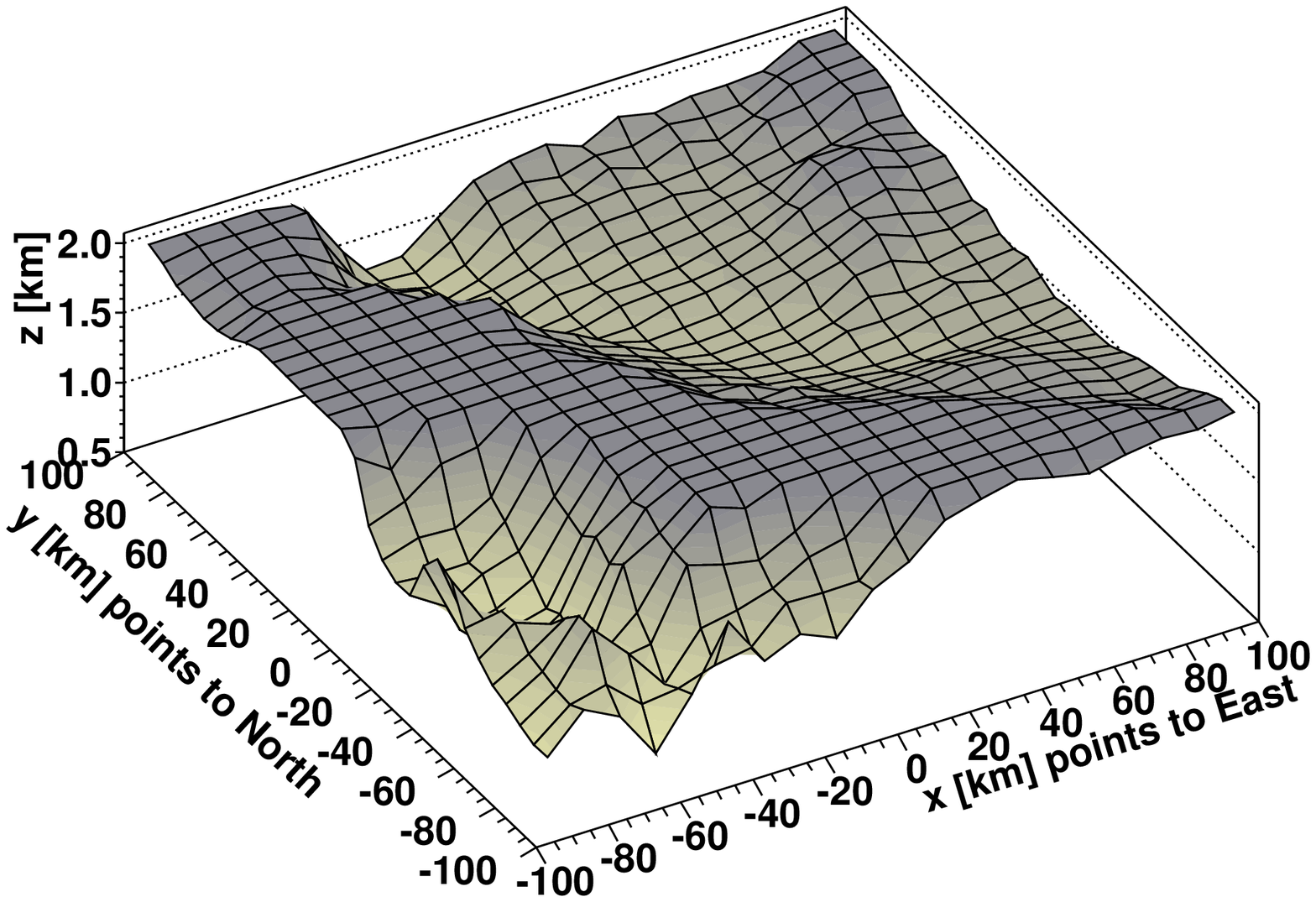} 

\includegraphics[width=\columnwidth,height=6.1cm]{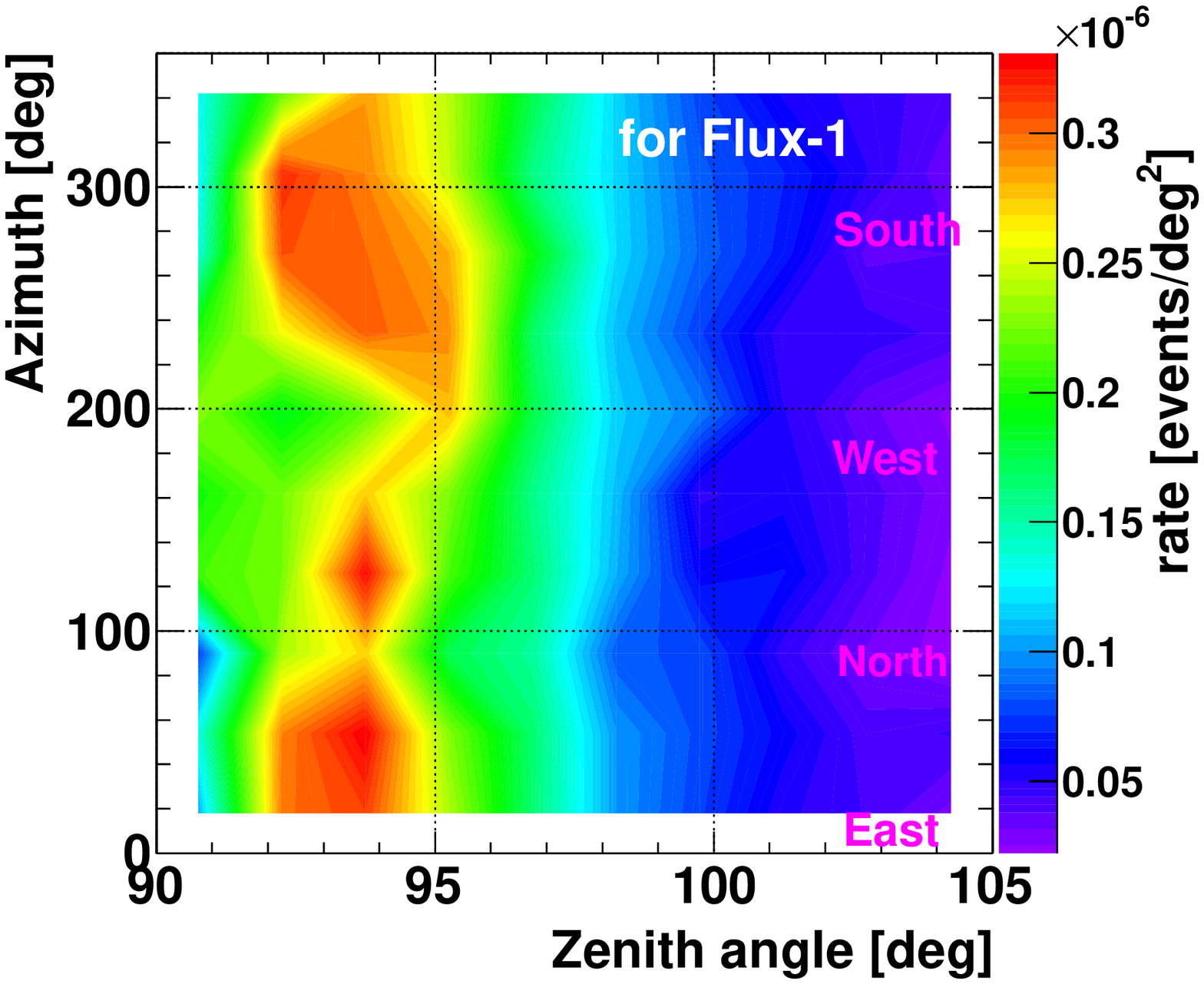} 
 \includegraphics[width=\columnwidth,height=6.1cm]{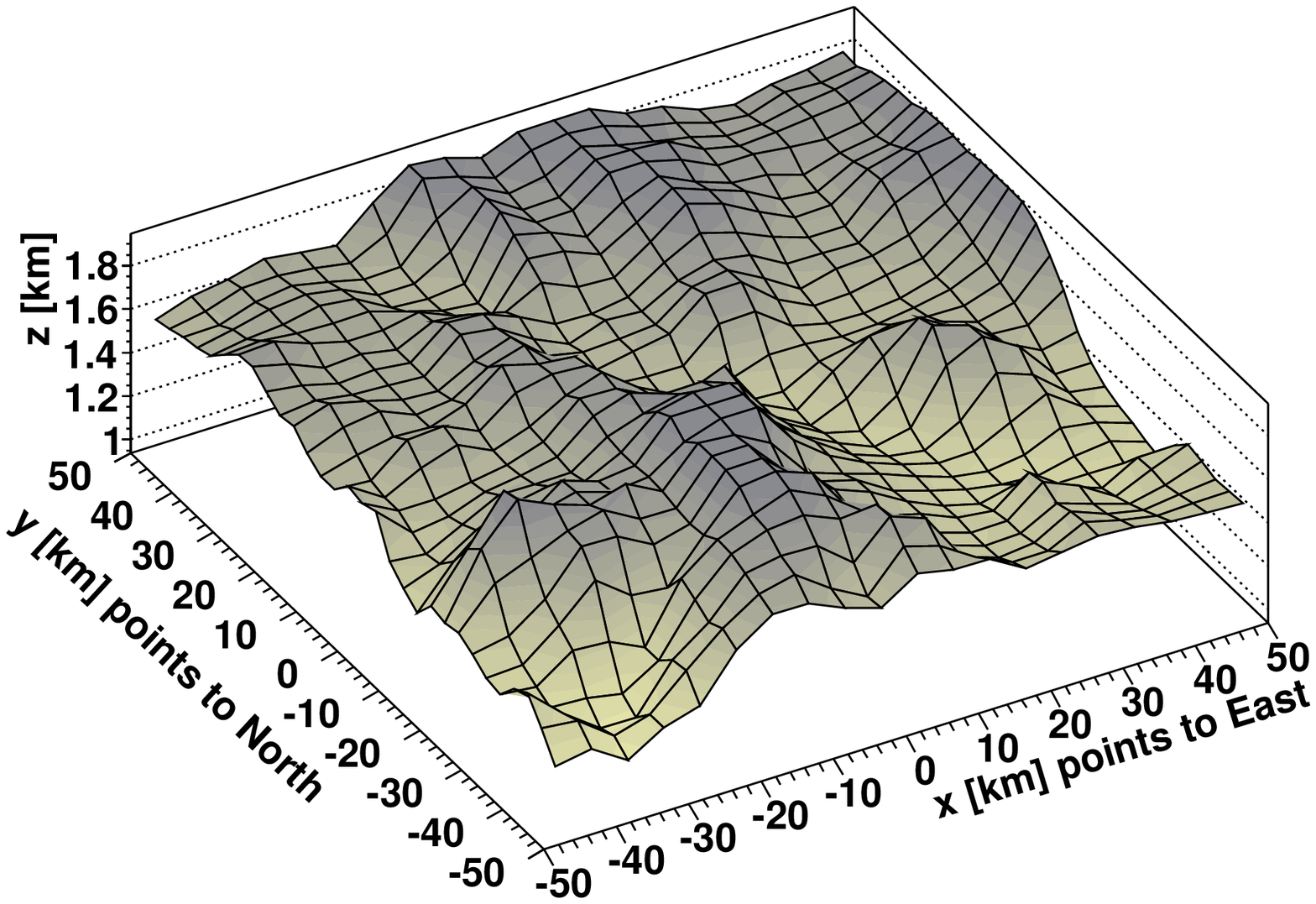} 
                                                    
  \end{center}                                                                                                               
  \caption{\label{fig::magic} {(Left column): Expected event rate calculated for Flux-1 as a function of azimuth and zenith for tau leptons with energies between $10^{16}$ eV and $10^{17}$ eV and a detector located at the VERITAS site (upper left), Meteor Crater (middle left) and Yavapai Ranch (bottom left); (Right column) Topography of considered sites according to CGIAR-CSI data: (Upper right) VERITAS telescopes (latitude $\phi_{\mathrm{VERITAS}}=31^{\circ} 40' 31''$ N, longitude $\lambda_{\mathrm{VERITAS}}=110^{\circ} 57' 07.77''$ W, height 1268 m a.s.l.). (Middle left ) the Meteor Crater site (latitude $\phi_{\mathrm{MeteorCrater}}=35^{\circ} 02' 46.7''$ N, longitude $\lambda_{\mathrm{MeteorCrater}}=111^{\circ} 02' 31.6''$ W, height 1677 m a.s.l.). (Bottom right panel) the Yavapai Ranch site  (latitude $\phi_{\mathrm{YavapaiRanch}}=35^{\circ} 08' 4.0''$ N, longitude $\lambda_{\mathrm{YavapaiRanch}}=112^{\circ} 51' 49.2''$ W, height 1658 m a.s.l.). The center of the maps corresponds to the position of  the considered site.}
 }                                                                                                                        
\end{figure*}   

\begin{table*}[bt!]
  \caption{\label{tab::rate222} {Expected event rates for Cherenkov detectors with trigger efficiencies of 10\% located at the different sites compared to that of IceCube (with a realistic efficiency). The values are calculated with the ALLM~\cite{allm} tau energy loss model and the GRV98lo~\cite{GRVlo} cross-section, with $f_{\mathrm{vis}}=100$\%, $\Delta \Omega=2\pi (\cos(90^{\circ})-\cos(105^{\circ}))=1.62$ and $\Delta T=3$ hours. The rates (in units $10^{-4}$)
 are calculated with the point source acceptances shown in Figure~\ref{fig::acccc}.}}
%\small
\begin{center}
\begin{tabular}{cccccccc}
\hline
\hline
 &Flux-1  &Flux-2&  Flux-3 & Flux-4 &Flux-5 \\
% &    &   &      &  & &\\
\hline
\hline
$N_{\mathrm{La Palma}}$  &2.8 & 1.5 & 0.86 &8.6 &2.6\\ 
$N_{\mathrm{VERITAS}}$   &8.2 & 3.9 & 1.6 &16 &6.9 \\
&    &   &      &  & &\\
$N_{\mathrm{Meteor Crater}}$  &9.0 & 4.2 & 1.7  &17 &7.5\\
$N_{\mathrm{Yavapai Ranch}}$  &7.7 & 3.7 & 1.5  &15 &6.5\\
&    &   &      &  & &\\
$N^{\mathrm{Northern \mbox{ } Sky}}_{\mathrm{IceCube}}$&  6.8&  2.5 &  0.46  & 4.6 & 8.8 \\
$N^{\mathrm{Southern \mbox{ } Sky}}_{\mathrm{IceCube}}$&  11.0&  3.2 &  0.76  & 7.6 & 8.8 \\
\hline
\hline
\end{tabular}
\end{center}
\end{table*}

In Fig.~\ref{fig::spectrum2}, a compilation of fluxes expected from AGN flares are shown. Flux-1 and Flux-2 are calculations for Feb 23, 2006 $\gamma$-ray flare of 3C 279~\cite{2009IJMPD}. Flux-3 and Flux-4 represent predictions for PKS~2155-304 in low-state and high-state, respectively~\cite{Becker2011269}. Flux-5 corresponds to a prediction for 3C~279 calculated in~\cite{PhysRevLett.87.221102}. 

The total observable rates (number of expected events) were calculated as $N=\Delta T \times \int_{E_{\mathrm{th}}}^{E_{\mathrm{max}}} A^{\mathrm{PS}}(E_{\nu_\tau})\times\Phi(E_{\nu_\tau})\times dE_{\nu_\tau}$, where $\Phi(E_{\nu_\tau})$ is the neutrino flux and $\Delta T$ the observation time (3 hours in Table~\ref{tab::rate222}).

Figure~\ref{fig::magic} shows the expected event rates for a detector with an average trigger efficiency of 10\% located at the VERITAS site, Meteor Crater and Yavapai Ranch, respectively, as a function of the incoming neutrino direction (defined by the zenith and azimuth angles). Correlations can be observed between the expected rate and the local topography. For the VERITAS site, the expected number of events from South-West and South-East is larger than from other directions due to the larger amount of matter encountered by incoming neutrinos from these directions. For tau lepton energies between $10^{16}$\,eV and $10^{17}$\,eV, the decay length is a few kilometers, hence, detectable events should mainly come from local hills not further away than about 50\,km. For Meteor Crater the largest mass distribution is seen in the  South-West, thus for this direction the calculated event rate is the largest. In case of the Yavapai Ranch site, the topography map shows many hills around the site leading to a rather complicated dependence of the event rate on azimuth and zenith. It is also worth to remark, that the presence of too many local hills close to the detector can in fact lead to a decrease of the up-going tau lepton flux and hence to a smaller event rate, as showers induced by up-going tau neutrinos are attenuated in the hills.

\section{Results} \label{sec:results}

\begin{figure}[ht]                                                                                                                
  \begin{center}                                                                                     
    \includegraphics[width=\columnwidth]{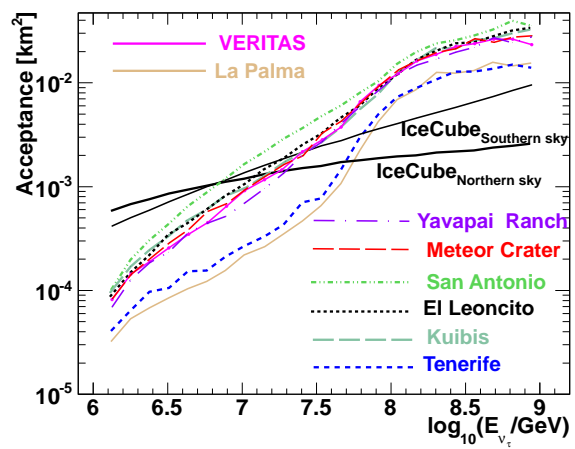}
  \end{center}
\vspace{-0.5cm}
  \caption{\label{fig::acccc} Acceptance for point source, $A^{\mathrm{PS}}(E_{\nu_\tau})$ to  earth-skimming tau  neutrinos as estimated for the La Palma site and a sample selection of few  future locations of Cherenkov instruments (with a trigger efficiency of 10\%) and  IceCube (as extracted from~\cite{IC-80-acc}). } 
\end{figure}

In Fig.~\ref{fig::acccc} we show as a function of the neutrino energy the estimated point source acceptance for the three studied sites in comparison to the La Palma site (MAGIC) and other possible locations of Cherenkov instruments.

Due to the lack of results from IceCube in the tau-neutrino channel, we use IceCube's muon neutrino acceptance \cite{IC-80-acc} for a sensitivity comparison. This is motivated by the fact that at Earth we expect an equal flavor flux from cosmic neutrino sources due to full mixing \cite{mixing}. In \cite{up-icecube,diff-icecube} it is also shown that for neutrino energies between 1\,PeV and 1000\,PeV, the muon-neutrino acceptance is only slightly larger than that for tau neutrinos.

The IceCube acceptance shows an increase for energies between $10^{6}$\,GeV and $10^{9}$\,GeV, and  is on average about $~2\times10^{-3}$ km$^{2}$. 
A potential detector with an average trigger efficiency of 10\% located at the VERITAS, Meteor Crater or Yavapai Ranch site can have an acceptance as large as a factor 10 greater than IceCube in the northern sky at energies larger than $\sim 5\times 10^{7}$\,GeV. For Flux-3 and Flux-4, this results in an event rate that is about a factor 3--4 larger than the rate calculated for IceCube in the northern sky assuming three hours of observation time (see Table~\ref{tab::rate222}). For neutrino fluxes covering the energy range below $\sim 5\times10^{7}$\,GeV (Flux-1, Flux-2, Flux-5), the number of expected events for these sites is comparable to that estimated for IceCube. 
Thus, Cherenkov telescopes pointing to the horizon, which is not their typical operating mode, could have a sensitivity comparable or even larger than that of neutrino telescopes such as IceCube in case of short neutrino flares (i.e.\ with a duration of about a few hours). For longer durations, the advantage of neutrino telescopes as full-sky 100\%-duty cycle instruments will be relevant. This can partially be compensated by Cherenkov telescopes with observations during nights with high clouds.

 Table~\ref{tab::rate222} also shows that in case of sites surrounded by mountains (VERITAS, Meteor Crater and Yavapai Ranch) event rates are at least a factor two higher than for sites without surrounding mountains (La Palma). 

The influence of systematic uncertainties is evaluated in terms of final event rates.
We  studied the influence on the expected event rate arising from uncertainties on the tau-lepton energy  loss and  different neutrino-nucleon cross-sections. The average energy loss of taus per distance travelled (unit depth $X$ in g\,cm$^{-2}$) can be described as $\left\langle dE/dX \right \rangle = \alpha(E) + \beta(E) E$. The factor $\alpha(E)$, which is nearly constant, is due to ionization. $\beta(E)$ is the sum of $e^+e^-$-pair production and bremsstrahlung (both well understood) and photonuclear scattering, which is not only the dominant contribution at high energies but at the same time subject to relatively large uncertainties. In this work, the factor $\beta_{\tau}$ is calculated using the following models describing contribution of photonuclear scattering: ALLM~\cite{allm}, BB/BS~\cite{bbbs} and CMKT~\cite{ckmt}, and different neutrino-nucleon cross-sections: GRV98lo~\cite{GRVlo}, CTEQ66c~\cite{cteq}, HP~\cite{hp}, ASSS~\cite{CooperSarkar:2007cv}, ASW~\cite{Albacete:2005ef}. Results are listed in~Table~\ref{tab::rate} for Flux-1 and Flux-3.

\begin{table}[h]
 \caption{\label{tab::rate} {Relative contributions to the systematic uncertainties on the up-going tau neutrino rate. As a reference value the expected event rate for the La Palma site  calculated for Flux-1 and Flux-3 was used.}}
\center 
\begin{tabular}{ccccc}
\hline
\hline
model &  PDF  & $\beta_{\tau}$ & sum     \\
\hline
\hline
Flux-1 
& $^{+14\%}_{-2\%}$    
& $^{+2\%}_{-7\%}$   
& $^{+14\%}_{-7\%}$ \\
& & &\\
Flux-3 
&  $^{+42\%}_{-7\%}$   
&  $^{+7\%}_{-14\%}$ 
& $^{+43\%}_{-16\%}$\\
\hline
\hline
\end{tabular}
\end{table}
%\vspace{-0.2cm}

Definitive conclusions on the capability of Cherenkov telescopes for detecting tau neutrinos, however,
require accurate neutrino-trigger and background simulations which go beyond the scope of this paper.

\begin{figure}[ht] 
  \begin{center}                                                                                     
    \includegraphics[width=\columnwidth]{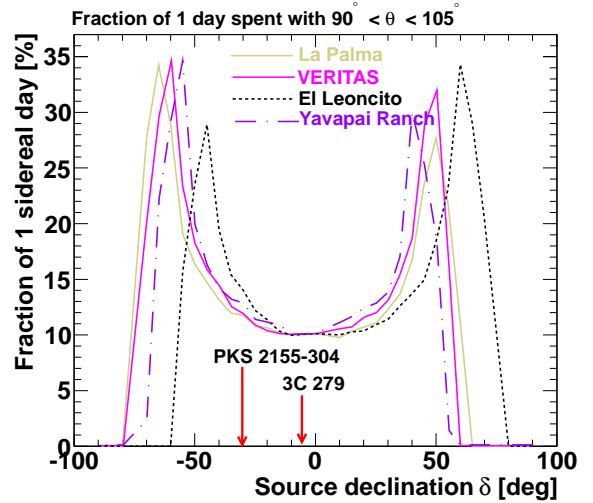}
  \end{center}
\vspace{-0.5cm}
  \caption{\label{fig::visibility} Fraction of a sidereal day in percent having a point-like source 
 at declination $\delta$ detectable with zenith angle $ 90^{\circ} <\theta<105^{\circ}$ i.e. for  
the  simulated zenith range  of  Earth-skimming (up-going) neutrinos .
 The fraction of time is plotted as a function of source declination  for a few selected sites.} 
\end{figure}

The acceptance/sensitivity to point sources also depends on the source declination ($\delta_s$), which defines the fraction of a day for which a source is visible in the sky at zenith angles between $90^\circ$ and $105^\circ$. Figure~\ref{fig::visibility} shows this fraction as a function of source declination for the different detector locations. As an example, 3C 279 ($\delta_s = - 5.8^{\circ}$) and PKS 2155-305 ($\delta_s = - 30.22^{\circ}$) are observed for $\sim 11$\% of a sidereal day (for about 2.4 hours per day), while for point sources with declination $\delta_s =-60^{\circ}$ the observation time is at least two times longer. The plot also shows that for the different detector locations the visibility has a quite similar behavior. In conclusion, even though the sensitivity window of the Earth-skimming analysis in zenith angle is small ($15^{\circ}$), it permits a point source survey in a broad range in declination angles spanning more than $100^{\circ}$ in the sky. 

% {\bf At the end, we would like to note that there has been no clear identification of tau neutrinos at high energies up to now i.e., no Double Bang event has been detected so far. Also, the recent IceCube results do not show any neutrino events at or above the Glashow resonance at 6.3 PeV~\cite{science_icecube}. This likely means that there is either a cutoff in the astrophysical neutrino flux below $\sim 6$ PeV or the neutrino spectrum is steeper than the usually assumed $E^{-2}$ spectrum. In both cases, the flux of tau neutrinos above 10 PeV where IACT win compared to IceCube is potentially rather low, thereby complicating the detection of Earth-skimming tau neutrinos with IACTs.}

\section{Summary}
%\vspace{-0.2cm}
In this paper, detailed Monte Carlo simulations have been employed to evaluate the sensitivity of air Cherenkov telescopes to high-energy  cosmic tau neutrinos in the PeV to EeV energy range coming from the horizon. The simulations use recent predictions for neutrino fluxes in AGN flares to calculate expected event rates at the VERITAS  site and two  future locations of Cherenkov instruments: Meteor Crater and Yavapai Ranch, taking into account the local topographic conditions.
The calculated neutrino rates are comparable to what has been estimated for the IceCube neutrino telescope assuming realistic observation times for Cherenkov telescopes of a few hours. For models with significant neutrino fluxes at energies above $\sim  5\times 10^{17}$\,eV, the sensitivity of Cherenkov telescopes actually surpasses that of IceCube. For the sites considered, the expected event rates are up to factor three higher compared to what is expected for the La Palma site (MAGIC).

%\vspace{-0.2cm}


\begin{thebibliography}{201}
\bibitem{uhecr} T.K Gaisser and T.Stanev, Astropart. Phys. 39-40 (2012) 120.
\bibitem{remnats} K. Koyama et al., Nature 378 (1995) 255.
\bibitem{Atoyan:2004pb} A.M. Atoyan, C.Dermer, New Astron. Rev. 48 (2004) 381.
\bibitem{PhysRevD.80.083008} A.Neronov, M.Ribordy, Phys. Rev. D 80 (2009) 083008.
\bibitem{Mucke2003593} A.Mucke, et al., Astropart. Phys. 18~(6) (2003) 593.
\bibitem{science_icecube}  M.G. Aartsen et al.,Science 342 (2013) 1242856. 
\bibitem{icecube_inter} I. Cholis and D. Hooper JCAP 06 (2013) 030; W. Winter, Phys. Rev. D 88, (2013) 083007;
O. Kalashev, et al., Phys.  Rev. Lett. 111 (2013) 041103; E. Roulet et al.; JCAP 01 (2013) 028;
 F.W. Stecker, Phys. Rev. D 88 (2013) 047301; H.N Ha et al., Phys. Rev.  D  87 (2013)  063011; 
K. Murase et al., Phys. Rev. D  88 (2013) 121301; L. Anchordoqui et al., Phys. Rev. D 89 (2014).083003; M.C. Gonzalez-Garcia et al., Astropar. Phys.  57  (2014) 39;	S. Razzaque, Phys. Rev. D, 88 (2013) 081302.

\bibitem{fargion0} D.~Fargion, astro-ph/9704205; astro-ph/0002453; The Astrophysical Journal 570 (2002) 909--925.
\bibitem{fargion1} D. Fargion et al., Report No HE.6.1.10; ICRC (1999),He 6.1.09, p.396-398.1999. 
Ed.D. Kieda, M. Salamon, B. Dingus(USA); D. Fargion et al., The Astrophysical Journal 613 (2004) 1285-1301.
\bibitem{zas}  H. Athar, G. Parente, E. Zas, Phys. Rev. D62 (2000) 093010.
\bibitem{Letessier:2001} A.~Letessier-Selvon, AIP Conf. Proc. 566 (2001) 157--171;
         X. Bertou et al., Astropart. Phys. 17 (2002) 183.
\bibitem{feng} J.L. Feng et al., Phys. Rev. Lett. 88 (2002) 161102.
\bibitem{tseng} J.J. Tseng et al. Phys. Rev. D 68 (2003) 063003. 
\bibitem{aramo} C. Aramo et al., Astropart. Phys. 23  (2005) 65-77.

\bibitem{learne} J.G. Learned et al., Astropart.Phys. 3 (1995) 267-274.
% \bibitem{oscarprl::2008} J.Abraham, et~al., Phys. Rev. Lett. 100 (2008) 211101;
%         J. Abraham, et~al., Phys. Rev. D, 79 (2009)  102001; P.Abreu, et~al. The Astrophysical Journal Letters, 755:L4 (2012); P. Abreu, et~al. Advances in High Energy Physics, 2013 (2013) 708680.
%\bibitem{up-icecube} R. Abbasi et. al, Phys.Rev. D86 (2012) 022005.

\bibitem{doi:10.1142/S0217732304014458} P.Yeh, et~al., Mod. Phys. Lett. A 19~(13n16) (2004) 1117;
Z.Cao, et~al., J. of Phys. G: Nucl. and Part. Phys. 31~(7) (2005) 571; J.Liu, et~al., J. Phys. G36 (2009) 075201.
\bibitem{Asaoka:2012em} Y.Asaoka, M.Sasaki, Astropart. Phys. 41 (2013) 7. 

\bibitem{magic} MAGIC {C}ollaboration: {http://http://magic.mppmu.mpg.de/}

\bibitem{veritas} VERITAS {C}ollaboration: {http://veritas.sao.arizona.edu/}

\bibitem{hess} HESS {C}ollaboration: {http://www.mpi-hd.mpg.de/hfm/HESS/pages/about/telescopes}
\bibitem{fargion} D.Fargion et. al., J. Phys. Conf. Ser. 110:062008, 2008; D. Fargion et al. 
Nucl. Instrum. Meth. A588 (2008) 146-150 arXiv:0710.3805 [astro-ph].
\bibitem{upgoing_magic} M. Gaug, C. Hsu, J.K. Becker, et al., Tau neutrino search with the MAGIC
telescope, in: International Cosmic Ray Conference, volume 3 of International
Cosmic Ray Conference, pp. 1273-1276.

\bibitem{cta} B.S.Acharya, et al., Astropart. Phys. 43 (2013) 6.

\bibitem{ouricrc} D. G\'ora and E. Bernardini, Proceedings of 33nd
International Cosmic Ray Conference - 2013 - Rio de Janeiro, Brazil,
arXiv:1308.0194.

\bibitem{gora:2007}
D.G\'ora, et al., Astropart. Phys. 26 (2007) 402.

\bibitem{dem}
{Consortium for Spatial Information (CGIAR-CSI)}.
{http://srtm.csi.cgiar.org/}

\bibitem{2009IJMPD}
A.{Reimer}, Int. Journ. of Mod. Phys. D 18 (2009) 1511.

\bibitem{Becker2011269}
J.K.Becker, et al., Nucl. Instr. and Meth. in Phys.
  Res. Sect. A: 630~(1) (2011) 269.
\bibitem{PhysRevLett.87.221102}
A.Atoyan, C.D.Dermer, Phys. Rev. Lett. 87 (2001) 221102.
\bibitem{IC-80-acc} J.A.Aguilar, astro-ph 1301.6504v1.
\bibitem{mixing} J. Beacom et al.,Phys. Rev. D68 (2003) 093005; H. Athar et al.,Phys. Rev.
D62 (2000) 103007.
\bibitem{up-icecube} R. Abbasi et. al, Phys.Rev. D86 (2012) 022005.
\bibitem{diff-icecube}  M.G. Aartsen, Phys. Rev. D88 (2013) 112008.

\bibitem{allm}
H.Abramowicz, A.Levy, hep-ph/9712415.
\bibitem{bbbs}
E.V.Bugaev, Y.V.Shlepin, Phys. Rev. D 67~(3) (2003) 034027.

\bibitem{ckmt}
A.Cappella, A.Kaidalov, C.Merino, J.Tran~Thanh, Phys. Lett. B 337 (1994)
  358.
\bibitem{GRVlo}
M.Gluck, E.Reya, A.Vogt, Eur. Phys. J. C5 (1998) 461.

\bibitem{cteq}
H.Lai, et~al., hep-ph/9903282.


\bibitem{hp}
A.Z.Gazizov, S.I.Yanush, Phys. Rev. D 65~(9) (2002) 093003.
\bibitem{CooperSarkar:2007cv}
A.Cooper-Sarkar, S.Sarkar, JHEP 0801 (2008) 075.

\bibitem{Albacete:2005ef}
J.L.Albacete, N.Armesto, J.G.Milhano, C.A.Salgado, U.A.Wiedemann,
 Eur. Phys. J. C43 (2005) 353.


\end{thebibliography}
\end{document}